\documentclass[preprint,amssymb,amsmath,amssymb,amsfonts,superscriptaddress,showpacs]{revtex4-1}

\usepackage{graphicx,graphics,pstricks,epsfig,wrapfig}

\begin{document}

\DeclareGraphicsExtensions{.pdf,.eps,.epsi,.jpg}

\title{}



\title{New antineutrino energy spectra predictions from the summation of beta decay branches of the fission products}

\author{M. Fallot}
\affiliation{SUBATECH, CNRS/IN2P3, Universit\'e de Nantes, Ecole des Mines de Nantes, F-44307 Nantes, France}

\author{S. Cormon}
\affiliation{SUBATECH, CNRS/IN2P3, Universit\'e de Nantes, Ecole des Mines de Nantes, F-44307 Nantes, France}

\author{M. Estienne}
\affiliation{SUBATECH, CNRS/IN2P3, Universit\'e de Nantes, Ecole des Mines de Nantes, F-44307 Nantes, France}

\author{A. Algora}
\affiliation{IFIC (CSIC-Univ. Valencia), Valencia, Spain}
\affiliation{Institute of Nuclear Research, Debrecen, Hungary}

\author{V.M. Bui}
\affiliation{SUBATECH, CNRS/IN2P3, Universit\'e de Nantes, Ecole des Mines de Nantes, F-44307 Nantes, France}

\author{A. Cucoanes}
\affiliation{SUBATECH, CNRS/IN2P3, Universit\'e de Nantes, Ecole des Mines de Nantes, F-44307 Nantes, France}

\author{M. Elnimr}
\affiliation{SUBATECH, CNRS/IN2P3, Universit\'e de Nantes, Ecole des Mines de Nantes, F-44307 Nantes, France}

\author{L. Giot}
\affiliation{SUBATECH, CNRS/IN2P3, Universit\'e de Nantes, Ecole des Mines de Nantes, F-44307 Nantes, France}

\author{D. Jordan}
\affiliation{IFIC (CSIC-Univ. Valencia), Valencia, Spain}

\author{J. Martino}
\affiliation{SUBATECH, CNRS/IN2P3, Universit\'e de Nantes, Ecole des Mines de Nantes, F-44307 Nantes, France}

\author{A. Onillon}
\affiliation{SUBATECH, CNRS/IN2P3, Universit\'e de Nantes, Ecole des Mines de Nantes, F-44307 Nantes, France}

\author{A. Porta}
\affiliation{SUBATECH, CNRS/IN2P3, Universit\'e de Nantes, Ecole des Mines de Nantes, F-44307 Nantes, France}

\author{G. Pronost}
\affiliation{SUBATECH, CNRS/IN2P3, Universit\'e de Nantes, Ecole des Mines de Nantes, F-44307 Nantes, France}

\author{A. Remoto}
\affiliation{SUBATECH, CNRS/IN2P3, Universit\'e de Nantes, Ecole des Mines de Nantes, F-44307 Nantes, France}

\author{J. L. Ta\'{\i}n}
\affiliation{IFIC (CSIC-Univ. Valencia), Valencia, Spain}

\author{F. Yermia}
\affiliation{SUBATECH, CNRS/IN2P3, Universit\'e de Nantes, Ecole des Mines de Nantes, F-44307 Nantes, France}

\author{A.-A. Zakari-Issoufou}
\affiliation{SUBATECH, CNRS/IN2P3, Universit\'e de Nantes, Ecole des Mines de Nantes, F-44307 Nantes, France}

\begin{abstract}
In this paper, we study the impact of the inclusion of the recently measured beta decay properties of the $^{102;104;105;106;107}$Tc, $^{105}$Mo, and $^{101}$Nb nuclei in an updated calculation of the antineutrino energy spectra of the four fissible isotopes 
$^{235, 238}$U, and $^{239,241}$Pu. These actinides are the main contributors to the fission processes in Pressurized Water Reactors. 
The beta feeding probabilities of the above-mentioned Tc, Mo and Nb isotopes have been found to play a major role in the $\gamma$ component of the decay heat of $^{239}$Pu, solving a large part of the $\gamma$ discrepancy in the 4 to 3000\,s range. They have been measured using the Total Absorption Technique (TAS), insensitive to the Pandemonium effect. 
The calculations are performed using the information available nowadays in the nuclear databases, summing all the contributions of the beta decay branches of the fission products. Our results provide a new prediction of the antineutrino energy spectra of $^{235}$U,  $^{239,241}$Pu and in particular of $^{238}$U for which no measurement has been published yet. We conclude that new TAS measurements are mandatory to improve the reliability of the predicted spectra.

\end{abstract}
\pacs{14.60.Pq, 23.40.-s,28.41.Ak, 28.41.-i,28.50.Hw}
\maketitle

Indications of a non-zero value of the neutrino oscillation mixing angle $\theta_{13}$ was obtained by combinations of fits\,\cite{fits} to KamLAND and solar data\,\cite{Solar}, MINOS\,\cite{Minos}, and T2K\,\cite{T2K}. Since then, new generation of reactor antineutrino disappearance experiments have come on line: Double Chooz\,\cite{DC}, Daya Bay\,\cite{DayaBay} and RENO\,\cite{RENO}, providing us with more precise measurements of $\theta_{13}$. As such, the quest for $\theta_{13}$ has triggered new studies, first initiated by Double Chooz, to compute reactor antineutrino energy spectra. 
These three new generation reactor experiments will have soon measurements of unprecedented precision reactor energy spectra thanks to their near detectors.  
Moreover, recent computations of reactor antineutrino energy spectra have shown evidence of the ``reactor anomaly"; a global deficit of the short baseline reactor experiments with respect to the new antineutrino flux predictions\,\cite{Mention}, motivating new searches for light sterile neutrinos\,\cite{WhitePaper}. 
In this context, it is important to evaluate reactor antineutrino energy spectra with independent methods.
The present letter wishes to contribute to this effort by providing new predictions of these spectra for $^{235}$U, $^{239}$Pu, $^{238}$U and $^{241}$Pu, ordered following the importance of their contribution to the fissions in Pressurized Water Reactors (PWRs). The new spectra are obtained by summation of the contributions of each fission product decay, and including the latest measurements of\,\cite{PRLAlgora}. Before discussing our results, we summarize the current situation concerning the predictions of Uranium and Plutonium isotope antineutrino energy spectra. 

Reactor antineutrinos arise from the $\beta^{-}$ decay of fission products created after the fissions of mainly $^{235}$U, $^{239, 241}$Pu and $^{238}$U.
K. Schreckenbach {\it et al.} have performed very precise measurements of the integral beta spectra of $^{235}$U, $^{239}$Pu and $^{241}$Pu at the Institut Laue Langevin, Grenoble, France\,\cite{SchreckU5-1, SchreckU5-2, SchreckU5Pu9,Schreckenbach} with energies up to 8\,MeV. In order to convert these beta spectra into antineutrino ones, the conservation of energy should be applied using each of the individual end-points of the beta branches. 
Since such information was not accessible by the authors, they fitted each integral beta spectrum with a set of 30 virtual beta branches. This conversion procedure induces a substantial shape uncertainty on the obtained antineutrino spectra. However, their measured spectra still remain the most precise ones, and the converted spectra were used in previous reactor neutrino experiments such as Bugey and CHOOZ\,\cite{bugey,chooz}.

In order to circumvent the conversion problem, O. Tengblad {\it et al.} have measured the individual beta spectra of 111 fission products\,\cite{tengblad} and combined these beta spectra with decay properties of 265 other fission products from nuclear databases. They have built the resulting beta spectra for $^{235, 238}$U, and $^{239}$Pu. Important errors were associated with the obtained global spectra and a notable disagreement was observed with previously mentioned results of\,\cite{SchreckU5-1, SchreckU5-2, SchreckU5Pu9,Schreckenbach}, that could probably be attributed to a lack of nuclear data. 

In\,\cite{Mueller}, these two methods were revisited. Firstly, the ``summation method" building the antineutrino energy spectrum through the fission product beta branches is explored again, thanks to the numerous information available nowadays in nuclear databases. Secondly, the conversion procedure mentioned in\,\cite{SchreckU5-1, SchreckU5-2, SchreckU5Pu9,Schreckenbach} is revisited, leading to new antineutrino spectra predictions and consequently to potentially new neutrino physics\,\cite{Mueller,Mention}. This result has been reinvestigated by P. Huber\,\cite{Huber}, confirming the conclusions of\,\cite{Mueller}.
A computation of the errors on the obtained antineutrino spectrum was also carried out in\,\cite{Mueller}, showing that, presently, the converted spectra from Schreckenbach {\it et al.} are the most precise ones. However, a normalisation change has been observed with respect to the previous predictions which shows that further investigation is needed. New decay measurements on selected fission products would bring new spectral predictions independent from the measurements of\,\cite{SchreckU5-1, SchreckU5-2, SchreckU5Pu9,Schreckenbach}, still unique since the 1980's. In addition, no measurement of the beta spectrum arising from the fission of $^{238}$U is available yet, making  the summation method the only method capable of predicting such a spectrum. It is also the only method able to predict the shape of the antineutrino energy spectra beyond 8\,MeV, the maximum energy given in the spectra of\,\cite{Mueller,Huber}. The binning of 250\,keV of the ILL measured spectra is also rather large, preventing the observation of structures that could arise from important contributors to the spectrum. 
We present below, new antineutrino energy spectra for the 4 main fissible isotopes built with the summation method, including  for the first time the latest TAS data measured by A. Algora {\it et al.}\,\cite{PRLAlgora} and using MURE simulation code.

The MCNP Utility for Reactor Evolution (MURE)\,\cite{MURE} is a precision, open-source code written in C++ that automates the preparation and computation of successive MCNP (Monte Carlo N-Particle\,\cite{MCNP}) calculations and solves the Bateman equations in between, for burnup or thermal-hydraulics purposes. 
The code was adapted to the needs of the simulation of antineutrino spectra and was used for the predictions already done in\,\cite{Mueller}, and for the calculation of the fission rates for the first results of the Double Chooz experiment\,\cite{DC}. MURE has been benchmarked with several codes (see for instance\,\cite{Takahama}).

In all antineutrino spectra presented below, the fission product concentrations are computed with MURE. The instantaneous thermal fission yields considered as an input of the MURE calculations are taken from\,\cite{JEFF}. The considered times of irradiation are the following: 12h for $^{235}$U, 1.5 days for $^{239,241}$Pu and 450 days for $^{238}$U.

The beta decay databases contain, when known, end-points, branching ratios, and spin/parities of beta transitions. A nucleus can be found in several databases. To avoid double counting and to select the preferred database for a given nucleus, we call the different databases following a priority order. For the presented spectra, the used databases are the following (ordered by priority): the Greenwood TAS dataset (29 nuclei)\,\cite{Greenwood}, the recently measured TAS dataset by A. Algora {\it et al.}\,\cite{PRLAlgora},  the experimental data measured by Tengblad {\it et al.}\,\cite{tengblad} (85 nuclei), experimental data from the evaluated nuclear databases JEFF31 (305, 345, 347 and 318 nuclei resp. for $^{235}$U, $^{239}$Pu, $^{241}$Pu and $^{238}$U)\,\cite{JEFF} and JENDL2000 (61,  62, 61, 58 nuclei resp.)\,\cite{JENDL}, ENSDF nuclei (87, 98, 101, 90 nuclei resp.)\,\cite{ENSDF}, Gross Theory spectra from JENDL (214, 215, 227, 221 nuclei resp.)\,\cite{GrossTheo} and the ``Q$_{\beta}$" approximation for the remaining unknown nuclei (22, 32, 38, 33 nuclei resp.). 
Indeed, for a non negligible amount of very neutron-rich fission products no data can be found in the databases quoted above. In this case, as a first milestone, the associated beta spectra could be built with 3 equiprobable allowed decay branches feeding the ground state, and excited states located at Q$_{\beta}$/3\,MeV and 2Q$_{\beta}$/3\,MeV above the ground state respectively. We call this approximation the ``Q$_{\beta}$" approximation. 

In\,\cite{Mueller}, it was noticed that the reconstructed spectra might suffer from an overestimation of the high energy part. This problem is due to the lack of data for exotic nuclei but also for a large part originates from the ``Pandemonium effect"\,\cite{Hardy77}. Indeed in decay databases some of the experimental information are biased due to the used detection apparatus.
When a daughter nucleus is fed in a high energy excited state, it may decay through a high energy gamma ray or through complex cascades of lower energy gamma rays. Germanium detectors were often used to measure the decay schemes of neutron-rich nuclei, but they suffer from a low intrinsic efficiency at high energy, giving a sizeable probability to miss high energy gamma transitions. Low energy gamma rays from complex decays can be blurred in the Compton fronts of the other decays.
Since the beta feeding probability is determined from the intensity balance feeding and de-exciting the levels, these features lead to an underestimation of high energy feeding and consequently an overestimation of the resulting beta spectra.

This is the reason why, in order to complete the amount of evaluated decay data already available, the JENDL database uses Gross Theory\,\cite{GrossTheo} to supply spectra for unknown nuclei and additional branches to correct a selected set of experimental data for nuclei assumed to suffer from the Pandemonium effect. 
Gross Theory spectra are only available for beta energy spectra. To obtain corresponding antineutrino spectra, we have used the beta branches characteristics given by Gross Theory but approximated the shape of the antineutrino spectrum of each branch by an allowed decay spectrum.

In total, these databases cover all the fission products present in the fission yield datafiles, i.e. 810, 874, 896 and 841 for $^{235}$U, $^{239}$Pu, $^{241}$Pu and $^{238}$U respectively. 
For each fission product from any of the databases, we reconstruct its antineutrino energy spectrum through the summation of its individual beta branches, using the prescription of\,\cite{Huber}, and taking into account the transition type, when given, up to the third forbidden unique transition. 
We have adopted the radiative correction formulae reported in\,\cite{Huber}, and the recommended weak magnetism value (0.67\,$\pm$\,0.26)\% m$_e$\,MeV$^{-1}$. Moreover, we have included in our spectra the weak interaction finite size correction and the screening correction, accounting for the screening of the nuclear charge by all the electrons in the atomic bound state. 
The overall effect of these two corrections is a $\le$\,1\% positive slope on the energy spectrum.

In this article, we present results of antineutrino spectra taking into consideration the latest published TAS data of the $^{102;104-107}$Tc, $^{105}$Mo, and $^{101}$Nb isotopes\,\cite{PRLAlgora}. The beta feeding of these nuclei were measured at the Ion-Guide Isotope Separator On-Line\,\cite{Jyvaskyla} facility of the University of Jyv\"{a}skyl\"{a}, coupling for the first time a Penning trap with a Total Absorption Spectrometer made with two large area NaI cristals of high efficiency. The TAS technique was previously applied by the same group in several experiments at GSI and ISOLDE\,\cite{ISOLDE}. 

In order to extract the beta feeding distributions of the nuclei of interest, the inverse problem was solved following the new reliable methods of analysis developed in\,\cite{PRLAlgora,Cano,Tain}. 
The $^{102;104-107}$Tc, $^{105}$Mo, and $^{101}$Nb nuclei belong to the list of nuclei suffering from the Pandemonium effect and contributing significantly to the $^{239}$Pu decay heat.  The results clearly show that the beta decay data of five of these nuclei were indeed seriously affected by the Pandemonium effect\,\cite{PRLAlgora}. It is thus essential to replace these nuclei by the new measurements in our computation of the antineutrino energy spectra of $^{235}$U, $^{239}$Pu, $^{241}$Pu and $^{238}$U.
The antineutrino spectra of the seven TAS nuclei could be built using each beta energy of the beta feeding distribution as an end-point of a beta-branch, the branching ratio being the feeding probability. Each decay is assumed to be an allowed transition. 
\begin{figure}[hbt!]
\includegraphics[scale=0.4]{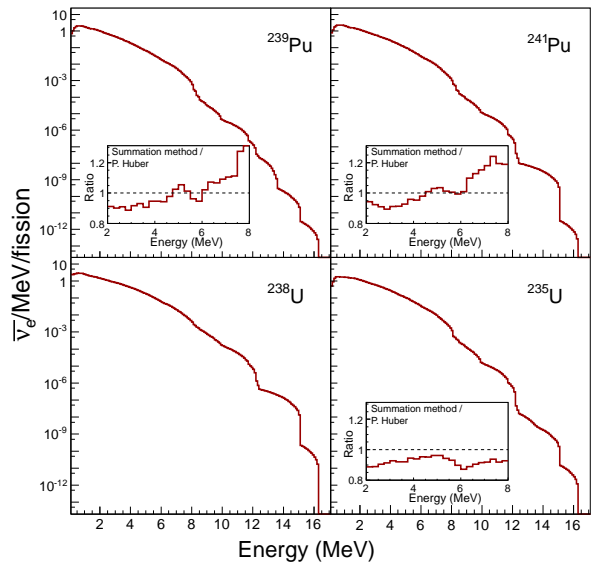}
\caption{\label{fig:Spectres} Reconstructed antineutrino energy spectra, including the latest TAS data from\,\cite{PRLAlgora}. In the inserts: the ratios of the spectra to the ones computed by Huber\,\cite{Huber}.}
\end{figure} 

\begin{figure}[hbt!]
\includegraphics[scale=0.4]{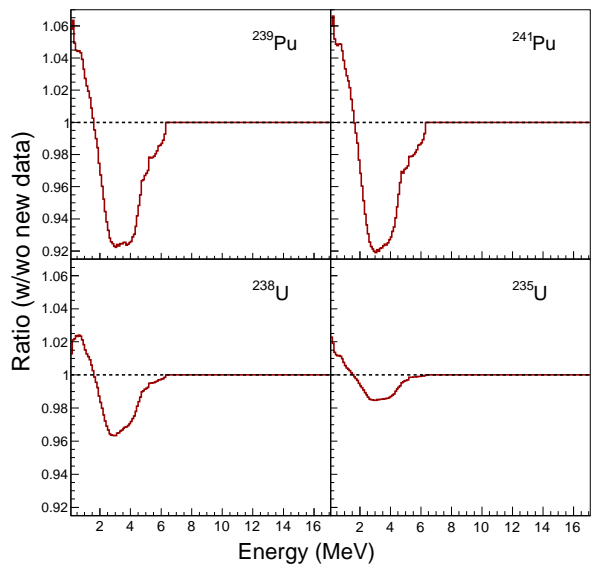}
\caption{\label{fig:Ratios} Ratios between 2 spectra built with the dataset described in the text over the same but the new TAS data. }
\end{figure} 
In Fig.\,\ref{fig:Spectres}, the new antineutrino energy spectra for $^{235, 238}$U, and $^{239,241}$Pu are displayed using 100\,keV bins. The spectra were obtained after the inclusion of the latter TAS data in our new computation. Note that below 1.8\,MeV, in a reactor core, actinides with small endpoints contribute and are not included in the presented spectra. Nevertheless the present results would be of interest for experiments using the inverse beta decay process to detect antineutrinos. In the inserts of Fig.\,\ref{fig:Spectres} the ratios of these spectra with the ones given in\,\cite{Huber}  are shown. The ratios are of the order of $\pm$\,10\% up to 7\,MeV for the isotopes. As for $^{235}$U, the ratio is constantly below one over the whole energy range, albeit within a 10\% envelope. We obtained similar ratios for the corresponding beta spectra over the ILL data\,\cite{SchreckU5-1, SchreckU5-2, SchreckU5Pu9,Schreckenbach}. Note the absence of insert in Fig.\,\ref{fig:Spectres} for $^{238}$U as no measurement has been published yet. The $^{238}$U antineutrino energy spectrum shown in Fig.\,\ref{fig:Spectres} provides thus a new prediction.
We have chosen a slightly different dataset from\,\cite{Mueller}, in which the data measured by Tengblad {\it et al.} have been given priority with respect to TAS data. Indeed, Tengblad {\it et al.}'s data may suffer from a systematic effect due to the calibration method. We thus have considered Greenwood's TAS data in priority, supplemented by the seven nuclei measured by Algora {\it et al.}. Nevertheless, the beta spectra measured by Tengblad {\it et al.} correct for a large number of Pandemonium nuclei, it is thus necessary to take these data into account preferentially to uncorrected datasets. Our new prediction lies in the $\pm$\,10\% range from the previous calculation in the energy range from 2 to 7\,MeV, and in the 20\% range between 7 and 8\,MeV.

The ratios of the new spectra with respect to the spectra obtained with the same set of data, but the latest TAS data for $^{102;104;105;106;107}$Tc, $^{105}$Mo, and $^{101}$Nb, are displayed in Fig.\,\ref{fig:Ratios}. The spectra of these seven nuclei were previously simulated using the JEFF experimental database.  
A noticeable deviation from unity is observed in the 0 to 6\,MeV energy range for the $^{239,241}$Pu energy spectra, reaching a 8\% decrease. As regards the $^{238}$U energy spectrum, the effect reaches a value of 3.5\% at 2.5-3\,MeV.
The discrepancy is smaller for the $^{235}$U isotope, 1.5\% at 2.5-3.5\,MeV, which is expected since these nuclei are a small contribution to the $^{235}$U spectrum. 

Considering the large impact of the replacement of these seven nuclei only, we have computed the resulting influence on detected antineutrino energy spectra by folding with the detection cross-section from\,\cite{CrossSection}. 
The resulting antineutrino flux corresponding to pure fissions of $^{235}$U, $^{239}$Pu, $^{241}$Pu, and $^{238}$U was found to be; 99.1\%, 94.53\%, 94.76\% and 98.09\%  relative to the flux obtained with the former dataset, respectively. 

To conclude, we have presented new predictions of the antineutrino energy spectra of the 4 main contributors to the fissions in a PWR, based on the summation of the beta decay branches of the fission products. The fission product concentrations were computed with the MURE code. The considered nuclear databases are, ordered by priority, the TAS data from Greenwood {\it et al.}\,\cite{Greenwood}, the data from Tengblad {\it et al.}\,\cite{tengblad}, the JEFF3.1 database\,\cite{JEFF}, the JENDL experimental database\,\cite{JENDL}, corrected with Gross Theory branches\,\cite{JENDL}, the Gross Theory spectra from JENDL\,\cite{GrossTheo}, and the ``Q$_{\beta}$" approximation for the remaining poorly known nuclei. The new predictions include the latest TAS measurements of seven nuclei, $^{102;104;105;106;107}$Tc, $^{105}$Mo, and $^{101}$Nb, free from the Pandemonium effect. The impact of these few nuclei reaches 8\% in the 2 to 4\,MeV energy range for the Plutonium isotopes, and 3.5\% for $^{238}$U. 
The effect is thus large and in an energy region of utmost importance for oscillation analysis of reactor neutrino experiments. Currently, the summation method is the only alternative to the antineutrino spectra converted from the integral beta spectra measured in the 1980's at ILL
The antineutrino spectra presented here would enable new generation reactor neutrino experiments to use a smaller binning and to complement their analysis at higher energies than 8\,MeV.  We should emphasize that presented here is a novel prediction of the antineutrino spectrum that arises from the fission of $^{238}$U, for which no measurements are available yet. Overall, the results presented in this paper show that Pandemonium nuclei play a major role in the estimate of the antineutrino spectra using the summation method, and that TAS measurements of these nuclei could allow to improve drastically the predictiveness of these spectra. The interest of these nuclear physics experiments is widened in the context of the reactor anomaly\,\cite{WhitePaper}, as independent evaluations of the reactor spectra could provide new constraints on the existence of light sterile neutrinos.

\section{Acknowledgements}
The authors would like to thank A. Sonzogni for very interesting discussions.
The authors from SUBATECH thank the challenge NEEDS and the GEDEPEON research groupment 
for their financial support, the Double Chooz collaboration and D. Lhuillier for providing ENSDF data under a user-friendly format.
The authors from IFIC thank the contribution of the FPA2011-24553 grant.

\end{document}